\documentclass[aps,nofootinbib,noshowpacs,preprintnumbers,amsmath,amssymb]{revtex4}
\pdfoutput=1

\usepackage{graphics}
\usepackage{graphicx}
\usepackage{dcolumn}
\usepackage{bm}
\usepackage{epsfig}
\usepackage{graphicx,color}
\usepackage{multirow}

\begin{document}


\title{Extraction of $\alpha_s$ using Borel-Laplace sum rules for tau decay data\footnote{Based on presentation at: {\it alphas-2022: Workshop on precision measurements of the QCD coupling constant\/}, January 31 - February 4, 2022, ECT* Trento, Italy; written for the Snowmass-2022 White Paper (The strong coupling constant: State of the art and the decade ahead)}}

\author{C\'esar Ayala$^a$}
\author{Gorazd Cveti\v{c}$^b$}
\author{Diego Teca$^b$}
\affiliation{$^a$Instituto de Alta Investigaci\'on, Universidad de Tarapac\'a, Casilla 7D, Arica, Chile}
\affiliation{$^b$Department of Physics, Universidad T{\'e}cnica Federico Santa Mar{\'\i}a (UTFSM), Casilla 110-V, Valpara{\'\i}so, Chile}

\date{\today}

\begin{abstract}
Double-pinched Borel-Laplace sum rules are applied to ALEPH $\tau$-decay data. For the leading-twist ($D=0$) Adler function a renormalon-motivated extension is used, and the 5-loop coefficient is taken to be $d_4=275 \pm 63$. Two $D=6$ terms appear in the truncated OPE ($D \leq 6$) to enable cancellation of the corresponding renormalon ambiguities. Two variants of the fixed order perturbation theory, and the inverse Borel transform, are applied to the evaluation of the $D=0$ contribution. Truncation index $N_t$ is fixed by the requirement of local insensitivity of the momenta $a^{(2,0)}$ and $a^{(2,1)}$ under variation of $N_t$. The averaged value of the coupling obtained is $\alpha_s(m_{\tau}^2)=0.3235^{+0.0138}_{-0.0126}$ [$\alpha_s(M_Z^2)=0.1191 \pm 0.0016$]. The theoretical uncertainties are significantly larger than the experimental ones. 
\end{abstract}

\maketitle

\vspace{0.5cm}

The sum rule corresponding to the application of the Cauchy theorem to a contour  integral containing the ($u$-$d$) quark $V+A$ correlator $\Pi(Q^2)$ ($Q^2 \equiv - q^2$) and a weight function $g(Q^2)$, $\oint_{C_1+C_2} d Q^2 g(Q^2) \Pi(Q^2) = 0$, gives the sum rule
\begin{equation}
\int_0^{\sigma_{\rm m}} d \sigma g(-\sigma) \omega_{\rm exp}(\sigma)  =
\frac{1}{2 \pi}   \int_{-\pi}^{\pi}
d \phi \; {\cal D}_{\rm th}(\sigma_{\rm m} e^{i \phi}) G(\sigma_{\rm m} e^{i \phi}) ,
\label{sr}
\end{equation}
where $\sigma_{\rm m} \equiv \sigma_{\rm max}$ is the maximal used energy in the data, and $\omega(\sigma)_{\rm exp}$ is the ALEPH-measured discontinuity (spectral) function of the $(V+A)$-channel polarisation function
\begin{equation}
\omega(\sigma) \equiv 2 \pi \; {\rm Im} \ \Pi(Q^2=-\sigma - i \epsilon) \ .
\label{om1}
\end{equation}
The function $g(Q^2)$ is the double-pinched Borel-Laplace weight function
\begin{equation}
g_{M^2}(Q^2) =  \left( 1 + \frac{Q^2}{\sigma_{\rm m}} \right)^2  \frac{1}{M^2} \exp \left( \frac{Q^2}{M^2} \right),
\label{g}
\end{equation}
$G(Q^2)$ is the integral of $g$
\begin{equation}
G(Q^2)= \int_{-\sigma_{\rm m}}^{Q^2} d Q^{'2} g(Q^{'2}),
\label{GQ2}
\end{equation}
and ${\cal D}_{\rm th}(Q^2)$ is the full Adler function ${\cal D}(Q^2) \equiv - 2 \pi^2 d \Pi(Q^2)/d \ln Q^2$, whose OPE truncated at dimension $D=6$ terms has the form
\begin{equation}
{\cal D}_{\rm th}(Q^2) = d(Q^2)_{D=0} + 1 + 4 \pi^2 \frac{\langle O_{4} \rangle}{ (Q^2)^2}  +   \frac{6 \pi^2}{ (Q^2)^3}  \left[ \frac{\langle O_{6}^{(2)} \rangle}{a(Q^2)} + \langle O_{6}^{(1)} \rangle \right] .
\label{DOPE}
\end{equation}
Here, $a(Q^2) \equiv \alpha_s(Q^2)/\pi$. The two terms of $D=6$ in the above OPE are needed to enable the cancellation of the corresponding $u=3$ IR renormalon ambiguities originating from the $D=0$ contribution $d(Q^2)_{D=0}$. The latter contribution has the perturbation expansion
\begin{equation}
d(Q^2)_{D=0, {\rm pt}}= d_0 a(\kappa Q^2) + d_1(\kappa) \; a(\kappa Q^2)^2 + \ldots + d_n(\kappa) \; a(\kappa Q^2)^{n+1} + \ldots, 
\label{dpt}
\end{equation}
where $\kappa \equiv \mu^2/Q^2$ is the renormalisation scale parameter ($0 < \kappa \lesssim 1$; usually $\kappa=1$), the first four terms ($d_0=1$; $d_1$, $d_2$, $d_3$) are exactly known \cite{Baikov:2008jh}, and for the coefficient $d_4$ [$\equiv d_4(\kappa)$ with $\kappa=1$ and $N_f=3$] we take the following values based on various specific estimates in the literature \cite{Kataev:1995vh,Baikov:2008jh,Boito:2018rwt,Beneke:2008ad,Cvetic:2018qxs}:
\begin{equation}
d_4 = 275 \pm 63 \ .
\label{d4est}
\end{equation}
The expansion of the Borel transform of $d(Q^2)_{D=0}$ is ${\cal B}[d](u;\kappa)_{\rm ser.} = \sum_{n \geq 0} d_n(\kappa) u^n/n!/\beta_0^n$.
The extension of $d(Q^2)_{D=0}$ beyond $\sim a^5$ is performed with a renormalon-motivated model \cite{Cvetic:2018qxs} in which the Borel transform is constructed first for an auxiliary quantity ${{\widetilde d}}(Q^2)$ of the Adler function \cite{Cvetic:2018qxs}, resulting in the Borel transform ${\cal B}[d](u)$ having terms $\sim 1/(2-u)^{{\widetilde {\gamma}}_2}$, $1/(3-u)^{{\widetilde {\gamma}}_3+1}$, $1/(3-u)^{{\widetilde {\gamma}}_3}$ and $1/(1+u)^{1+{\overline {\gamma}}_1}$, and similar terms with lesser powers, where ${\widetilde {\gamma}}_p=1 + p \beta_1/\beta_0^2$ (${\overline {\gamma}}_p=1 - p \beta_1/\beta_0^2)$, and $\beta_0=(11 - 2 N_f/3)/4$ and $\beta_1=(1/16)(102 - 38 N_f/3)$ are the first two $\beta$-function coefficients ($N_f=3$).\footnote{In our ansatz \cite{Cvetic:2018qxs} and notation, the effective one-loop $D=6$ anomalous dimensions  $-\gamma_{O_6}^{(1)}/\beta_0$ (appearing beside ${\widetilde {\gamma}}_3$ in the mentioned powers ${\widetilde {\gamma}}_3-\gamma_{O_6}^{(1)}/\beta_0$) were taken to be large-$\beta_0$, i.e., $-\gamma_{O_6}^{(1)}/\beta_0=1, 0$. The work \cite{Boito:2015joa} implies that these quantities can be evaluated beyond large-$\beta_0$, resulting in a decreasing sequence of nine numbers $-\gamma_{O_6}^{(1)}/\beta_0 \approx  -0.197; -0.247; \ldots$. It remains an open question how to extend the renormalon-motivated \cite{Cvetic:2018qxs} model to include these results.}
This extension gives, for the choice $d_4=275.$, the coefficients of the expansion (\ref{dpt}): $d_5=3159.5$; $d_6=16136.$; $d_7=3.4079 \times 10^5$; $d_8=3.7816 \times 10^5$; $d_9=6.9944 \times 10^7$; $d_{10}=-5.8309 \times 10^8$; etc. 

The cancellation of the IR renormalon ambiguity requires: (i) for $u=2$ IR renormalon term ${\cal B}[d](u) \sim 1/(2-u)^{{\widetilde {\gamma}}_2}$, the $D=4$ OPE term of the Adler function to be of the form $1/(Q^2)^2$; (ii) for the $u=3$ IR renormalon term ${\cal B}[d](u) \sim 1/(3-u)^{{\widetilde {\gamma}}_3}$ to be of the form $1/(Q^2)^3$; (iii) and for the $u=3$ IR renormalon term ${\cal B}[d](u) \sim 1/(3-u)^{{\widetilde {\gamma}}_3+1}$ to be of the form $1/(Q^2)^3/a(Q^2)$. These three terms ($D=4, 6$) are taken into account in the OPE (\ref{DOPE}).

The $D=0$ contribution $d(Q^2)_{D=0}$ to the Adler function in the sum rule contour integral (\ref{sr}) is evaluated in three different ways. We apply two variants of fixed order perturbation theory (FOPT). In the first variant the powers of $a(\kappa \sigma_{\rm m} e^{i \phi})^n$ are expressed as truncated Taylor series in powers of $a(\kappa \sigma_{\rm m})$ (FO). In the second variant  $d(Q^2)_{D=0}$ is expressed as the sum of the logarithmic derivatives ${\widetilde a}_n(Q^2)$ [$\propto (d/ d \ln Q^2)^{n-1} a(Q^2)$], and then ${\widetilde a}_n(\kappa \sigma_{\rm m} e^{i \phi})$ are expressed as truncated Taylor series of ${\widetilde a}_k(\kappa \sigma_{\rm m})$ (${\widetilde {\rm FO}}$). The third way of evaluation is the use of the inverse Borel transformation of $d(Q^2)_{D=0}$, where the Borel integral is evaluated with the Principal Value (PV) prescription; in the integrand, the Borel transform ${\cal B}[d](u)$ is taken as as a series consisting of the mentioned (renormalon-related) inverse powers $\sim (p-u)^k/(p-u)^{\gamma}$ ($k=0,1,\ldots$), where the series is truncated; this truncation requires for $d(\sigma_{\rm m} e^{i \phi})_{D=0}$ introduction of an additional correction polynomial  $\delta d(\sigma_{\rm m} e^{i \phi})_{D=0}^{[N_t]}$ in powers of $a(Q^2)$. In all the three methods, a truncation index $N_t$ is involved, i.e., only the terms up to the power $a^{N_t}$ (or ${\widetilde a}_{N_t}$ in ${\widetilde {\rm FO}}$) are taken into account.

\begin{figure}[htb] 
\centering\includegraphics[width=90mm]{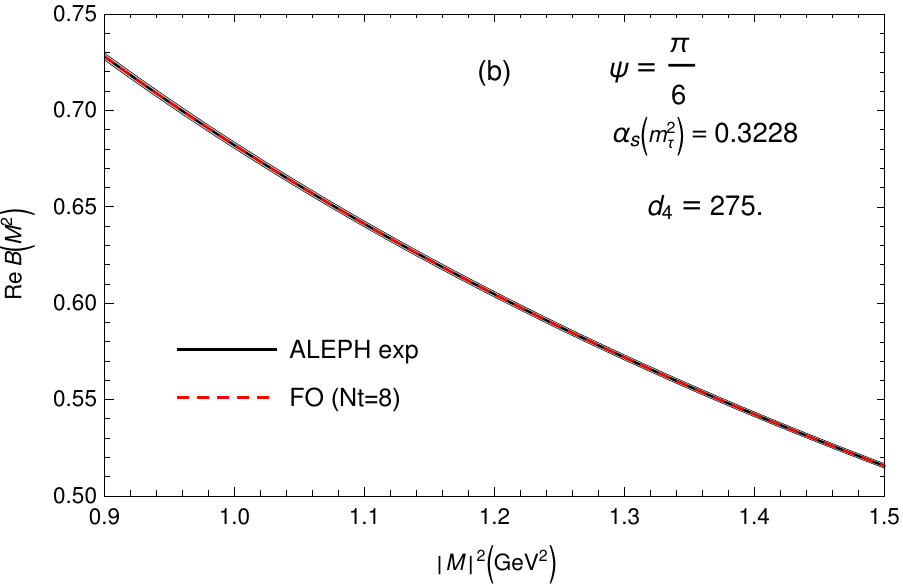}
\caption{(coloured online) The values of ${\rm Re} B(M^2;\sigma_{\rm m})$ along the ray $M^2=|M^2| \exp( i \Psi)$ with $\Psi=\pi/6$. The narrow grey band are the experimental predictions. The red dashed line inside the band is the result of the FOPT global fit with truncation index $N_t=8$. Similar fitting curves are obtained for the rays with $\Psi=0$ and $\Psi=\pi/4$.}
\label{fig:FigPsiPi6}
\end{figure}

We apply the Laplace-Borel sum rules, with the weight function (\ref{g}), to the ALEPH $V+A$ data with $\sigma_{\rm max} (\equiv \sigma_{\rm m})$ $=2.8 \ {\rm GeV}^2$ (i.e., the last two bins are excluded due to large uncertainties). In the sum rule (\ref{sr}), this gives on both sides the Borel-Laplace sum rule quantity $B(M^2;\sigma_{\rm m})$. In practice, the rule is applied to the real parts only, ${\rm Re} B_{\rm exp}(M^2;\sigma_{\rm m}) = {\rm Re} B_{\rm th}(M^2;\sigma_{\rm m})$, and for the scale parameters $M^2$ along rays in the first quadrant: $M^2=|M^2| \exp(i \Psi)$ with $0 \leq \Psi < \pi/2$. We minimise (with respect to $\alpha_s$, $\langle O_4 \rangle$,  $\langle O_6^{(1)} \rangle$ and $\langle O_6^{(2)} \rangle$) the  following sum of squares:
\begin{equation}
\chi^2 = \sum_{\alpha=0}^n \left( \frac{ {\rm Re} B_{\rm th}(M^2_{\alpha};\sigma_{\rm m}) - {\rm Re} B_{\rm exp}(M^2_{\alpha};\sigma_{\rm m}) }{\delta_B(M^2_{\alpha})} \right)^2 ,
\label{xi2} \end{equation}
where $\{ M_{\alpha}^2 \}$ was taken as a dense set of points along the chosen rays with $\Psi=0, \pi/6, \pi/4$  and $0.9 \ {\rm GeV}^2 \leq |M_{\alpha}|^2 \leq 1.5 \ {\rm GeV}^2$. We chose 11 equidistant points along each of the three rays, and the series (\ref{xi2}) thus contains 33 terms (the fit results remain practically unchanged when the number of points is increased). In the sum (\ref{xi2}), the quantities $\delta_B(M^2_{\alpha})$ are the experimental standard deviations of ${\rm Re} B_{\rm exp}(M^2_{\alpha};\sigma_{\rm m})$, with the ALEPH covariance matrix for the $(V+A)$-channel taken into account (cf.~App.~C of \cite{Ayala:2017tco} for more explanation). For each evaluation method (FO, ${\widetilde {\rm FO}}$, PV) and for each chosen truncation index $N_t$, the fit procedure gives us results, and the fit is usually of good quality ($\chi^2 \lesssim 10^{-3}$), cf.~Fig.~\ref{fig:FigPsiPi6}.

The truncation index $N_t$ is then fixed by considering the first two double-pinched momenta $a^{(2,0)}(\sigma_{\rm m})$ and $a^{(2.1)}(\sigma_{\rm m})$\footnote{The weight functions for double-pinched momenta $a^{(2,n)}$ are: $g^{(2,n)}(Q^2) = ((n+3)/(n+1)) (1/\sigma_{\rm m}) (1 + Q^2/\sigma_{\rm m})^2 \sum_{k = 0}^{n} (k + 1)(-1)^k (Q^2/\sigma_{\rm m})^k$. The obtained values of $a^{(2,0)}(\sigma_{\rm m})$ and $a^{(2.1)}(\sigma_{\rm m})$ are well within the experimental band.}
and requiring local stability of their values under the variation of $N_t$.
The resulting extracted values of the coupling are
\begin{eqnarray}
\alpha_s(m_{\tau}^2)^{\rm (FO)} & = & 0.3228 \pm 0.0003({\rm exp})^{-0.0026}_{+0.0070}(\kappa)^{-0.0103}_{+0.0079}(d_4)^{+0.0081}_{-0.0057}(N_t)
\nonumber\\
& = &  0.3228^{+0.0134}_{-0.0121} \approx 0.323^{+0.013}_{-0.012} \qquad (N_t=8^{+2}_{-3}).
\label{BLalFO} \\
\alpha_s(m_{\tau}^2)^{\rm ({\widetilde {\rm FO}})} & = & 0.3209 \pm 0.0003({\rm exp})^{-0.0038}_{+0.0201}(\kappa)^{-0.0039}_{+0.0047}(d_4)^{+0.0293}_{-0.0084}(N_t)
\nonumber\\
& = &  0.3209^{+0.0359}_{-0.0100} \approx 0.321^{+0.036}_{-0.010} \qquad (N_t=5 \pm 2).
\label{BLaltFO} \\
\alpha_s(m_{\tau}^2)^{\rm (PV)} & = & 0.3269 \pm 0.0003({\rm exp})^{+0.0007}_{+0.0102}(\kappa)^{-0.0064}_{+0.0155}(d_4)^{+0.0092}_{-0.0006}(N_t)^{+0.0167}_{-0.0067}({\rm amb})
\nonumber \\
& = &  0.3269^{+0.0266}_{-0.0093} \approx 0.327^{+0.027}_{-0.009} \qquad (N_t=8^{+2}_{-3}).
\label{BLalPV} \\
\alpha_s(m_{\tau}^2)^{\rm (CI)} & = & 0.3488 \pm 0.0005({\rm exp})^{+0.0078}_{+0.0004}(\kappa) \pm 0.0000(d_4)^{-0.0027}_{+0.0119}(N_t)
\nonumber \\
& = &  0.3488^{+0.0142}_{-0.0028} \approx 0.349^{+0.014}_{-0.003}  \qquad (N_t=4^{+2}_{-1}).
\label{BLalCI}
\end{eqnarray}
The uncertainties were presented as separate terms. The variation of the renormalisation scale parameter $\kappa \equiv \mu^2/Q^2$ was taken in the range $2/3 \leq \kappa \leq 2$ ($\kappa=1$ for the central values). The truncation index is $N_t=8,5,8$ for the central cases of FO, ${\widetilde {\rm FO}}$ and PV.

The (truncated) Contour improved perturbation theory (CIPT) results were also included in the above results, for comparison. However, the truncated CIPT approach for $d(Q^2)_{D=0}$ evaluation appears to require a different type of OPE in the $D>0$ part of the contributions, because the renormalon structure and the related renormalon ambiguities are not reflected in the truncated CIPT series \cite{Hoang:2021nlz}.
We thus include only FO, ${\widetilde {\rm FO}}$ and PV results in the average
\begin{eqnarray}
\alpha_s(m_{\tau}^2) &=& 0.3235^{+0.0138}_{-0.0126} \qquad ({\rm FO}+{\widetilde {\rm FO}}+{\rm PV}) 
\nonumber \\
\Rightarrow \;
\alpha_s(M_Z^2) &=& 0.1191 \pm 0.0016.  
\label{3av} \end{eqnarray} 
In Table \ref{tab:rescomp} we compare these results with some other results in the literature.
\begin{table}
  \caption{\footnotesize The values of $\alpha_s(m_{\tau}^2)$, extracted by various groups applying sum rules and various methods to the ALEPH $\tau$-decay data.}
 \label{tab:rescomp}
\centering
 \begin{ruledtabular}
  \begin{tabular}{l|l|lll|l}
group &  sum rule & FO & CI & PV & average \\
\hline
Baikov et al.,2008~\cite{Baikov:2008jh} & $a^{(2,1)}=r_{\tau}$ & $0.322 \pm 0.020$ & $0.342 \pm 0.011$ & --- & $0.332 \pm 0.016$ \\
Beneke\&Jamin, 2008~\cite{Beneke:2008ad} & $a^{(2,1)}=r_{\tau}$ &  $0.320^{+0.012}_{-0.007}$ & --- & $0.316 \pm 0.006$ & $0.318 \pm 0.006$ \\
Caprini, 2020 \cite{Caprini:2020lff} & $a^{(2,1)}=r_{\tau}$ &  --- & --- & $0.314 \pm 0.006$ &  $0.314 \pm 0.006$ \\
Davier et al., 2013~\cite{Davier:2013sfa} & $a^{(i,j)}$ & $0.324$ & $0.341 \pm 0.008$ & --- & $0.332 \pm 0.012$ \\
Pich\&R.S\'anchez, 2016~\cite{Pich:2016bdg}   &  $a^{(i,j)}$     & $0.320 \pm 0.012$ &  $0.335 \pm 0.013$ & --- & $0.328 \pm 0.013$  \\
Boito et al., 2014~\cite{Boito:2014sta} & DV in $a^{(i,j)}$ & $0.296 \pm 0.010$ & $0.310 \pm 0.014$ & --- & $0.303 \pm 0.012$ \\
our prev. work, 2021~\cite{Ayala:2021mwc}  & BL ($O_6, O_8$) & $0.308 \pm 0.007$ & ---   & $0.316^{+0.008}_{-0.006}$ & $0.312 \pm 0.007$ \\
\hline
this work, 2022 (also~\cite{Ayala:2021yct})  & BL  ($O_6^{(1)}, O_6^{(2)}$) & $0.323^{+0.013}_{-0.012}$(FO) &  ---   & $0.327^{+0.027}_{-0.009}$ & $0.324 \pm 0.013$
\\
  &  &  $0.321^{+0.021}_{-0.030}$(${\widetilde {\rm FO}}$) & & & 
  \end{tabular}
\end{ruledtabular}
\end{table}

The results Eq.~(\ref{3av}) can get significantly affected when the assumptions or methods are changed. For example, if we chose, instead of the central value $d_4=275.$, the upper upper bound $d_4=338.$ of Eq.~(\ref{d4est}) as the central value, the results would decrease somewhat, to $\alpha_s(m_{\tau}^2) \approx 0.320 \pm 0.015$
[$\alpha_s(M_Z^2) \approx$  $0.1187^{+0.0016}_{-0.0019}$], i.e., $\delta \alpha_s(m^2_{\tau}) \approx -0.004$.

If we took, instead of the two mentioned $D=6$ terms in the OPE, the simple $D=6$ and $D=8$ OPE terms [$\sim 1/(Q^2)^3$ and $\sim 1/(Q^2)^4$], the central value would decrease by about $\delta \alpha_s(m^2_{\tau}) \approx -0.008$. In our previous work \cite{Ayala:2021mwc} we used the OPE with simple $D=6,8$ terms, and took for $d_4$ higher values $d_4=338 \pm 63$ than here Eq.~(\ref{d4est}).

 If we took $N_t=5$ in all three methods (i.e., no extension of Adler function beyond $d_4 a^5$), then the central value of $\alpha_s(m_{\tau}^2)$ in FO changes from $0.3288$ to $0.3171$, and in PV from $0.3269$ to $0.3277$ $\Rightarrow$ for the average of the three methods the central value changes from $0.3235$ to $0.3219$ [$\alpha_s(M_Z^2)$ from $0.1191$ to $0.1189$], i.e., $\delta \alpha_s(m^2_{\tau}) = -0.0016 \approx -0.002$, small.

  According to the results (\ref{BLalFO})-(\ref{BLalPV}), Borel-Laplace sum rules indicate that the theoretical uncertainties dominate over the experimental ones. Part of these theoretical uncertainties would be reduced by: 1.) the calculation of the five-loop Adler function coefficient $d_4$; 2.) the use of the more complicated structure of the $D=6$ OPE terms \cite{Boito:2015joa} and the corresponding terms in the $D=0$ $u=3$ IR renormalon structure; 3.) the use of a variant of the QCD coupling $a(Q^2)$ without the Landau singularities in the $D=0$ contribution, because this would allow for the resummation to all orders (no truncation) of the renormalon-motivated contribution $D=0$ and would eliminate the renormalisation scale ambiguity ($\kappa$). The high precision ALEPH determination of the $\tau$ spectral function represents an important source of data for understanding better the behaviour of QCD at the limit between the perturbative and nonperturbative regimes.      



\end{document}